\documentclass[a4paper,fleqn]{cas-dc}
\usepackage[numbers,sort&compress]{natbib}
\usepackage[dvipsnames]{xcolor}
\usepackage{graphicx, hyperref, listings, xspace}
\hypersetup{colorlinks=true, linkcolor=BurntOrange, citecolor=BurntOrange, filecolor=BurntOrange, urlcolor=BurntOrange}
\newcommand{\ie}{\textit{i.e.}\xspace}

\newcommand{\mgamc}{{\sc MG5\_aMC}\xspace}
\newcommand{\py}{{\sc Pythia~8}\xspace}
\def\madanalysis{{\sc MadAnalysis~5}\xspace}
\def\fastjet{{\sc FastJet}\xspace}
\def\met{E_T^\mathrm{miss}}

\DeclareUnicodeCharacter{2212}{-}

\begin{document}
\title[mode=title]{Prospects for toponium formation at the LHC in the single-lepton mode}
\let\printorcid\relax   
\author[1]{Benjamin Fuks}
\author[2]{Kaoru Hagiwara}
\author[2, 3]{Kai Ma}
\author[1]{Léandre Munoz-Aillaud}
\author[4]{Ya-Juan Zheng}

\address[1]{Laboratoire de Physique Théorique et Hautes Énergies (LPTHE), UMR 7589, Sorbonne Université et CNRS, 4 place Jussieu, 75252 Paris Cedex 05, France}
\address[2]{KEK Theory Center, Tsukuba, Ibaraki 305-0801, Japan}
\address[3]{Faculty of Science, Xi'an University of Architecture and Technology, Xi'an, 710055, China}
\address[4]{International College, The University of Osaka, Toyonaka, Osaka 560-0043, Japan}

\shortauthors{B.~Fuks {\it et~al.}}
\shorttitle{Prospects for toponium formation at the LHC in the single-lepton mode}

\begin{abstract}
  We investigate the formation of toponium in the single-leptonic final state at the LHC. Our study builds on our recently proposed framework that incorporates the associated non-perturbative effects into Monte Carlo simulations through the Green's function of the non-relativistic QCD Hamiltonian and the re-weighting of hard-scattering matrix elements. This allows us to perform a phenomenological analysis that demonstrates that a statistically significant excess from toponium formation could already be accessible in Run~2 data. Moreover, our results highlight observables that provide handles for signal characterisation and establish the single-leptonic channel as a competitive and complementary avenue for the ongoing exploration of toponium signatures at colliders.
\end{abstract}

\maketitle

\vspace*{-9.75cm}
  \noindent {\small\texttt{KEK-TH-2751}}
\vspace*{8.75cm}


\section{Introduction}
The study of top-antitop production at the LHC plays a central role in testing the Standard Model and probing potential new physics. The top quark is indeed produced copiously in highly energetic proton-proton collisions, with cross sections measured with increasing precision across different final states~\cite{ATLAS:2022aof, CMS:2023qyl, ATLAS:2023slx, ATLAS:2024kxj}. Furthermore, the dynamics of top-antitop pairs near threshold provide a unique laboratory to explore the strong interaction in the presence of heavy quarks. While bound states of heavy quark-antiquark pairs such as charmonium and bottomonium are well established, the extremely short lifetime of the top quark prevents the formation of narrow toponium resonances~\cite{Fadin:1987wz, Fadin:1990wx, Hagiwara:2008df, Sumino:2010bv}. Nevertheless, non-relativistic QCD (NRQCD) predicts the existence of remnant bound-state effects that could leave measurable imprints in the threshold region of top-antitop production. Their observation would thus offer an unprecedented window on the interplay between perturbative and non-perturbative QCD at the electroweak scale. 

Recently, both the ATLAS and CMS collaborations have independently reported excesses in the dileptonic $t\bar t$ channel close to threshold, and these excesses have been found consistent with expectations from toponium formation~\cite{ATLAS:2023fsd, CMS:2024ybg, CMS:2024pts, CMS:2024zkc, CMS:2025kzt, CMS:2025dzq, ATLAS:2025kvb}. These observations further confirm earlier indications based on partial Run~2 data~\cite{ATLAS:2019hau, ATLAS:2023gsl}, while our study~\cite{Fuks:2021xje} has helped to revive theoretical and phenomenological interest in bound-state effects in top physics at the LHC~\cite{Maltoni:2024tul, Aguilar-Saavedra:2024mnm, Llanes-Estrada:2024phk, Fuks:2024yjj, Garzelli:2024uhe, Francener:2025tor, Nason:2025hix, Fuks:2025sxu, Bai:2025buy, Shao:2025dzw}. More recently, we proposed a strategy to incorporate non-perturbative NRQCD contributions from toponium formation into the Monte Carlo simulations commonly used in LHC analyses~\cite{Fuks:2024yjj}. The key ingredient of this framework is the family of Green's functions of the non-relativistic QCD Hamiltonian~\cite{Fadin:1987wz, Fadin:1990wx,  Strassler:1990nw, Jezabek:1992np, Sumino:1992ai, Hagiwara:2008df}, which provide the bound-state corrections in a form that can be used to re-weight conventional $t\bar t$ production and decay matrix elements. In this way, collider-level simulations including parton showering, hadronisation, detector response and event reconstruction become feasible. At present, the publicly available implementation employs only the colour-singlet S-wave Green's function, but the structure of the framework readily accommodates the consistent inclusion of higher-order and higher-multiplicity components as they become available. Complementarily, fixed-order predictions including higher-order QCD corrections and threshold resummation have been achieved for specific observables such as the $t\bar t$ invariant-mass spectrum~\cite{Kiyo:2008bv, Ju:2020otc, Garzelli:2024uhe, Nason:2025hix}. Taken together, these developments provide both the theoretical tools and the motivation to explore toponium signatures in LHC data.

In this work, we generalise the approach proposed in~\cite{Fuks:2024yjj} to design a strategy for incorporating toponium effects into Monte Carlo simulations of $t\bar t$ production near threshold, focusing on final states in which one top quark decays semi-leptonically and the other hadronically. This single-leptonic channel provides an ideal balance between statistical yield and kinematic reconstruction power. On the one hand, it benefits from the larger branching ratio relative to the dileptonic mode while still featuring a clean charged-lepton signature for triggering and background suppression. On the other hand, the presence of a single neutrino allows for full event reconstruction through a kinematic fit, which potentially provides a powerful access to observables sensitive to near-threshold dynamics. From a theoretical standpoint, however, conventional Monte Carlo event generators currently lack a consistent treatment of non-perturbative bound-state effects in this channel. Embedding toponium corrections directly into simulations of this final states is therefore essential for the exploration of top-antitop threshold effects at the LHC, as well as for assessing their potential impact on precision top-quark measurements and new-physics searches.

In the remainder of this letter, we present our strategy for incorporating toponium effects into Monte Carlo simulations relevant for top-antitop production and decay in the single-lepton mode in section~\ref{sec:technics}, and we apply it in section~\ref{sec:pheno} to a dedicated phenomenological analysis. This study not only illustrates the feasibility of probing toponium formation with existing LHC Run~2 data, but also identifies by means of event simulations including reconstruction and selection effects a set of observables that might be particularly sensitive to it. We summarise and conclude in section~\ref{sec:conclusion}.

\section{Technicalities}\label{sec:technics}

In this report, we generalise the approach proposed in~\cite{Fuks:2024yjj} to incorporate toponium effects into Monte Carlo simulations for hadron colliders, using the \mgamc event generator~\cite{Alwall:2014hca} (version 3.5.7) and its capabilities for the automatic generation of leading-order matrix elements. While our earlier works were dedicated to final states with two leptons, \ie the process $gg\to (t \bar{t})_1 \to b\ell^+\nu_\ell\, \bar{b}\ell^{\prime-}\bar{\nu}_{\ell'}$ with the top-antitop quark pair being in a colour-singlet state, we adapt the method to address final states with a single charged lepton $\ell$ (taken as either a muon or an electron for phenomenological purposes). The two hard-scattering processes considered are
\begin{equation}\label{eq:processes}
  g g \!\to\! (t\bar{t})_1 \!\to\! b \ell^+ \nu_\ell \bar{b} q\bar{q}'\ \ \text{and}\ \
  g g \!\to\! (t\bar{t})_1 \!\to\! b q\bar{q}' \bar{b} \ell^{\prime-} \bar{\nu}_\ell'\,.
\end{equation} 
In such a toponium signal definition, only diagrams containing two resonant top propagators are included. Single-resonant and non-resonant contributions leading to the same six-fermion final state are thus expected to be treated as part of what we refer to in this study as the background, \ie the conventional simulation of the full $t\bar t$ final state in which Coulombic gluon effects near threshold are not included. Furthermore, no on-shell constraint is imposed on the intermediate top and antitop quarks, while the six final-state fermions are required to be on-shell. These processes hence explicitly include the top and antitop decays, which is essential to open the relevant phase-space region below the $t\bar{t}$ production threshold and to correctly embed all spin correlations among the final-state particles. A corresponding working directory can be generated in \mgamc by typing
\begin{lstlisting}
  generate g g > t t$\sim$ > w+ b w- b$\sim$,\
      w+ > l+ vl, w- > j j
  add process g g > t t$\sim$ > w+ b w- b$\sim$,\ 
      w+ > j j, w- > l- vl$\sim$
  output toponium_1l
\end{lstlisting}

This set of commands enforces that all generated diagrams involve two (possibly off-shell) internal top quark propagators and a six-body final state. The two intermediate top quarks can then exchange Coulomb gluons relevant for toponium formation, and these effects are incorporated by modifying the source code from the \lstinline{toponium_1l} directory generated by \mgamc. To this aim, the implementation requires two key modifications. First, the intermediate top-antitop quark pair is restricted to be in a colour-singlet configuration in order to isolate the subset of diagrams that can experience the attractive QCD potential relevant for bound-state formation. Second, the squared matrix elements $|\mathcal{M}|^2$ associated with the processes~\eqref{eq:processes} must be re-weighted by the ratio of the Green's functions of the non-relativistic QCD Hamiltonian and the free Hamiltonian, $\widetilde{G}(E; p^*)/\widetilde{G}_0(E; p^*)$, where $E$ represents the toponium binding energy and $p^*$ is the recoil momentum of the top quark in the toponium rest frame.\footnote{The Green's functions also depend on the top mass and width, which are kept implicit for brevity.} This factor effectively resums the Coulombic QCD interactions between the heavy top quarks and encodes the modifications of the cross section near threshold due to bound-state dynamics. The leading-order squared matrix element generated by \mgamc is thus replaced by
\begin{equation}\label{eq:reweighting}
  |\mathcal{M}|^2 \to \bigg|\mathcal{M}\, \frac{\widetilde{G}(E; p^*)}{\widetilde{G}_0(E; p^*)}\bigg|^2\,,
\end{equation}
where we implicitly include both contributing processes~\eqref{eq:processes}. We emphasise that this re-weighting procedure employs the colour-singlet Green's function from the NRQCD Hamiltonian, but it does not correspond to a full NRQCD factorisation of the process; the underlying matrix elements indeed remain those generated in the standard collinear factorisation. Whereas the free Green's function is the standard one, 
\begin{equation}
  \widetilde{G}_0(E; p^*) = \frac{1}{(E+i\Gamma_t) - \frac{p^{*2}}{m_t}}\,,
\end{equation}
with $m_t$ and $\Gamma_t$ denoting the top mass and width respectively, the interacting Green's function $\widetilde{G}(E; p^*)$ can be obtained by solving the Lippmann-Schwinger equation with the QCD potential~\cite{Jezabek:1992np, Hagiwara:2016rdv, Hagiwara:2017ban}. We refer to section~2 of~\cite{Fuks:2024yjj} for comprehensive details.

These adjustments are implemented by modifying four of the \textsc{Fortran} files generated by \mgamc, namely the two occurrences of the pair of files \lstinline{matrix1_orig.f} and \mbox{\lstinline{driver.f}} located in the folders \lstinline{P1_gg_wpbwmbx_wp_lvl_wm_qq} and \lstinline{P2_gg_wpbwmbx_wp_qqwm_lvl} of the \lstinline{SubProcesses} directory. The modifications follow exactly the procedure detailed in section~3 of~\cite{Fuks:2024yjj}, with one modification specific to the single-leptonic setup: the colour matrix must be multiplied by a global factor of three arising from the colour structure of the final-state quark-antiquark pair produced in a $W$ decay. The initial colour data,
\begin{lstlisting}[mathescape=false]
  C     COLOR DATA
  C     
    DATA (CF(I,1),I=1,2)/
   $ 1.600000000000000D+01,
   $ -2.000000000000000D+00/
  C     1 T(1,2,5,8) T(6,7)
    DATA (CF(I,2),I=1,2)/
   $ -2.000000000000000D+00,
   $  1.600000000000000D+01/
  C     1 T(2,1,5,8) T(6,7)
  C     ----------
\end{lstlisting}
must therefore be replaced by
\begin{lstlisting}[mathescape=false]
  C     COLOR DATA
  C     
    DATA (CF(I,1),I=1,2)/
   $ 2.000000000000000D+00,
   $ 2.000000000000000D+00/
  C     1 T(1,2,5,8) T(6,7)
    DATA (CF(I,2),I=1,2)/
   $ 2.000000000000000D+00,
   $ 2.000000000000000D+00/
  C     1 T(2,1,5,8) T(6,7)
  C     ----------
\end{lstlisting}
Finally, to prevent \mgamc from optimising the calculation and overriding the modifications made, the automatic recycling of helicity amplitudes must be disabled by adding the following line at the end of the \lstinline{run_card.dat} configuration file located in the \lstinline{Cards} sub-folder of the \lstinline{toponium_1l} directory:
\begin{lstlisting}
  False = hel_recycling
\end{lstlisting}

At this stage, hard-scattering event generation with the \mgamc platform follows the usual procedure, initiated from the working directory by typing in a shell the command
\begin{lstlisting}
  ./bin/generate_events
\end{lstlisting}
Correctly matching the resulting events with parton showering requires preventing the colour-singlet intermediate virtual top and antitop quarks from radiating gluons. The ground toponium state has a size of about $(20~\mathrm{GeV})^{-1}$, which indeed forbids QCD emissions with energies below 20~GeV from the bound system. In this way, only initial-state radiation and final-state radiation from the five lightest quarks and gluons are allowed. To enforce this, the produced event file is modified by updating the colour-flow information of each event such that both the initial gluon pair and the final state are in a colour-singlet configuration. For this purpose, we follow the procedure described in the last part of section~3 of~\cite{Fuks:2024yjj} and introduce a phantom intermediate $Z'$ resonance carrying the four-momentum of the toponium state. This purely technical trick ensures that parton showering as handled with \py~\cite{Bierlich:2022pfr} both preserves the invariant mass of the top-antitop system and prevents it from radiating gluons. In addition, we modify the event files to explicitly re-introduce intermediate particles that \mgamc\ omits from the event record when they are too far off shell, so that the full decay chain is consistently preserved in each event. We remind that by default, \mgamc indeed only writes intermediate particles to the event record if their invariant mass lies within $n=15$ widths of the pole mass, the value of $n$ being controlled from the \lstinline{bwcutoff} parameter of the \lstinline{run_card.dat} configuration file. 

After these modifications, a typical single-leptonic toponium event in the LHE format would thus resemble  
\begin{lstlisting}
  <event>
   13 2 +1.51e+00 1.76e+02 7.55e-03 1.08e-01
       21 -1    0    0  504  506 ...
       21 -1    0    0  506  504 ...
       32  2    1    2    0    0 ...
       6   2    3    3  505    0 ...
       24  2    4    4    0    0 ...
      -6   2    3    3    0  505 ...
      -24  2    6    6    0    0 ...
       4   1    5    5  501    0 ...
      -3   1    5    5    0  501 ...
       5   1    4    4  505    0 ...
       13  1    7    7    0    0 ...
      -14  1    7    7    0    0 ...
      -5   1    6    6    0  505 ...
  </event>
\end{lstlisting}
where the momentum information has been replaced with ellipses and we have limited the floating-point numbers in the first line of the event to three digits for brevity. Following the standard LHE format~\cite{Boos:2001cv, Alwall:2006yp}, each of the subsequent lines corresponds to a particle. Their first entry indicates the particle identifier according to the Particle Data Group numbering scheme and their second entry refers to the initial-state, intermediate-state or final-state nature of the particle. The third and fourth entries provide information on the parent of each particle, while the fifth and sixth entries encode the colour and anticolour information.

To achieve this in a straightforward manner, we updated the \lstinline{Python} script previously used in our work on toponium di-leptonic decays, and we adapted it to the single-leptonic case. This script will prepare the generated hard-scattering events for parton-shower simulations by updating the colour-flow information for the bound-state configuration, introducing the phantom $Z'$ carrying the toponium four-momentum and adding any missing top or antitop from the event record. The resulting file \lstinline{reprocess_1l.py} is available from \href{https://github.com/BFuks/toponium.git}{https://github.com/BFuks/toponium.git} and should be executed within any of the \lstinline{Events/run_xx} sub-folders generated by \mgamc. Once these modifications are complete, parton showering can be simulated from the working directory by typing in a shell the command
\begin{lstlisting}
  ./bin/madevent 
\end{lstlisting}
This launches the command-line interface of \mgamc, from which the showering step can be initiated with
\begin{lstlisting}
  pythia8 run_xx
\end{lstlisting}
with \lstinline{run_xx} denoting the relevant run name.

\section{Single-leptonic signatures of toponium at the LHC}\label{sec:pheno}
We now turn to our phenomenological analysis aiming to assess the impact of toponium formation on the singly-leptonic-decaying $t\bar{t}$ final state. In the following, we use the term `\textit{signal}' to refer solely to the subset of $t\bar t$ events in which the intermediate top-antitop pair is in a colour-singlet state and is re-weighted using the NRQCD Green's function, thereby incorporating bound-state effects.  Conversely, the term `\textit{background}' denotes the standard perturbative prediction for the same six-fermion final state, including single-resonant and non-resonant contributions but without any bound-state corrections. We emphasise that this terminology is used only to distinguish these two components within our Monte Carlo implementation and is not intended to suggest the presence of a new physical process beyond ordinary $t\bar t$ production. Moreover, at the level of the leading-order matrix elements considered here, the Green's function re-weighting of the signal adds Coulombic corrections which are absent from the background, so the two contributions are complementary and do not overlap; potential double counting would indeed only appear if higher-order corrections were included in the background, which is beyond the scope of this work.

We employ \mgamc to generate both standard top-antitop background events and toponium events (following the procedure detailed in the previous section). We consider 140~fb$^{-1}$ of proton-proton collisions at the LHC with a centre-of-mass energy $\sqrt{s}=13$~TeV and convolute the associated $2\to 6$ leading-order matrix elements with the CT18NLO parton distribution functions~\cite{Hou:2019efy} as provided through \lstinline{LHAPDF}~\cite{Buckley:2014ana}. Two independent samples of 100,000 signal and 1,000,000 background hard-scattering events are hence produced for a top quark mass and width fixed to $m_t = 173$~GeV and $\Gamma_t = 1.49$~GeV, using a tree-level Coulomb-like potential for the calculation of the NRQCD Green's function. The events are subsequently showered and hadronised with \py~\cite{Bierlich:2022pfr}, and then reconstructed and analysed within \madanalysis~\cite{Conte:2018vmg, Araz:2020lnp} using the anti-$k_T$ jet algorithm~\cite{Cacciari:2008gp} with a radius parameter of $R = 0.4$ as implemented in \fastjet~\cite{Cacciari:2011ma}. Moreover, we normalise the leading-order background event sample to a total cross section computed at next-to-next-to-leading order with next-to-next-to-leading logarithmic threshold resummation (NNLO+NNLL)~\cite{Czakon:2013goa}, using a value of 243~pb. On the other hand, the toponium signal is normalised to the cross section of 1.51~pb obtained directly from our Monte Carlo simulations including Green's function re-weighting.

To enhance sensitivity to near-threshold bound-state effects, we preselect events featuring one reconstructed electron or muon with a transverse momentum $p_T>10$~GeV and pseudo-rapidity $|\eta|<2.5$, as well as two $b$-jets and two light jets with $p_T>25$~GeV and $|\eta|<2.5$. In addition, lepton candidates lying within $\Delta R<0.4$ of any jet are rejected so that only isolated leptons are retained, and the selected events must feature a missing transverse energy $\met>30$~GeV.

A practical advantage of the single-leptonic channel compared to the dileptonic case is that the event reconstruction is simpler thanks to the presence of only one final-state neutrino. To illustrate this, we perform a full reconstruction of the selected events. We first identify the pair of light jets whose invariant mass is closest to the $W$-boson mass \mbox{$m_W=80.419$~GeV}. The neutrino longitudinal momentum $p_z^\nu$ is then determined by combining the missing transverse momentum with the reconstructed lepton and requiring the resulting invariant mass to match the value of $m_W$. This condition yields a quadratic equation in $p_z^\nu$. If no real solution exists, we enforce a single solution by setting the square root term to zero. On the other hand, if two real solutions are found, both are tested. For each case, the leptonic and hadronic top candidates are reconstructed by pairing the two $b$-jets with either the lepton-neutrino system or the dijet system. The $b$-jet assignment and neutrino momentum choice that minimise the deviation of the reconstructed top masses from $m_t$ is then retained. Subsequently, we continue the event selection procedure by restricting the reconstructed invariant mass of the $t\bar t$ system to be below 350~GeV.

\begin{figure}
  \centering
  \includegraphics[width=\linewidth]{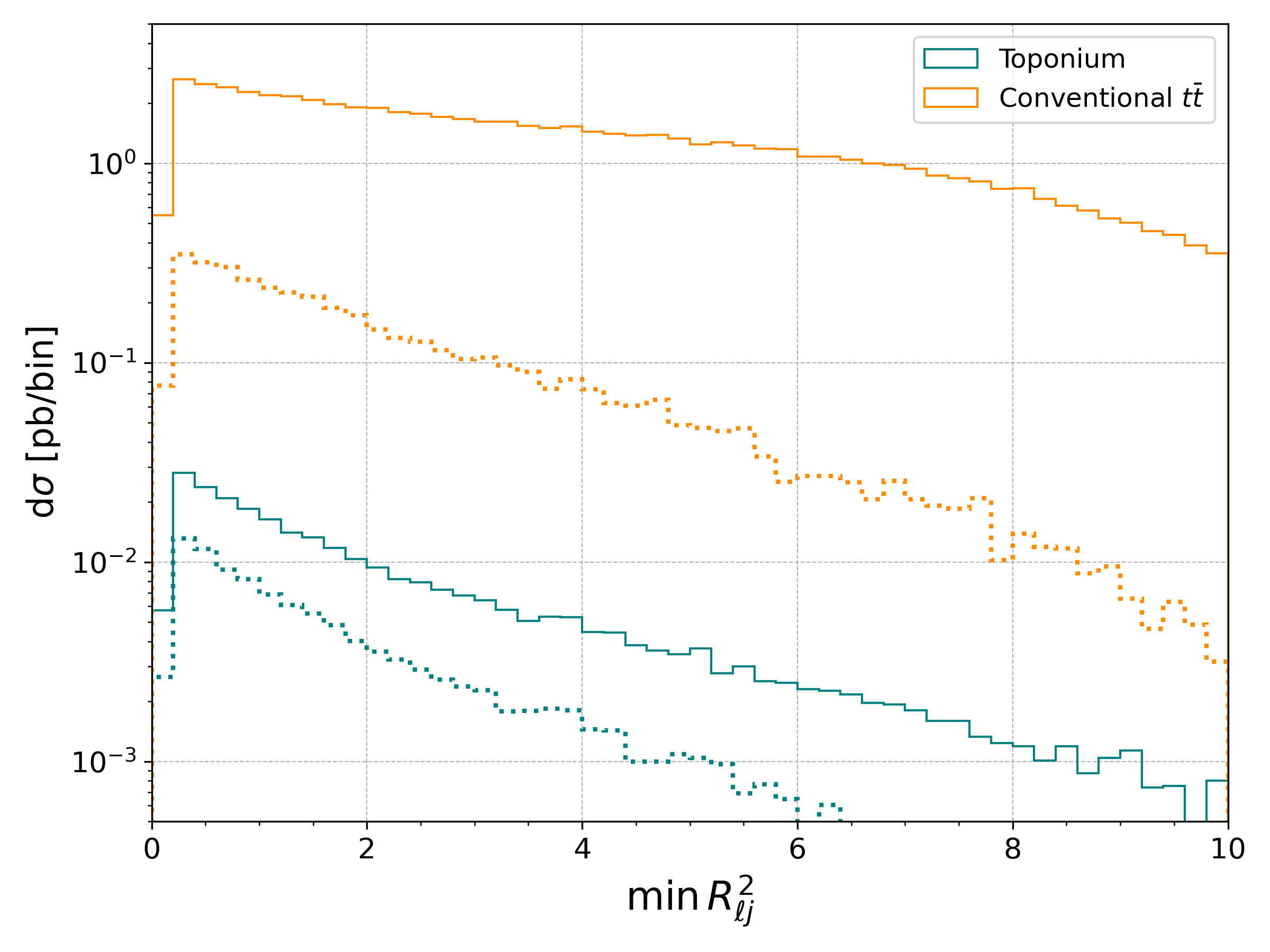}
  \caption{Differential cross section for toponium (teal) and conventional top-antitop (orange) production in proton-proton collisions at $\sqrt{s}=13$~TeV in the single-leptonic channel, shown as a function of the minimum angular separation between the lepton and the two leading light jets. The predictions include a fiducial selection requiring exactly one lepton, two light jets, two $b$-jets and $\met > 30$~GeV. In addition, we present the results both when the reconstructed invariant mass of the top-antitop system is constrained to be below 350~GeV (dashed) or not (solid).\label{fig:R2lj}}
\end{figure}

Motivated by our earlier studies~\cite{Fuks:2021xje, Fuks:2024yjj} relying on the properties of the $t\bar t$ production spin polarisation density matrix~\cite{Hagiwara:2017ban} convoluted with the top and antitop decay density matrices, we focus on events in which the separation in the transverse plane between the lepton originating from the semi-leptonic top decay and the down-type quark initiated jet from the fully hadronic top decay is small. Such a choice originates directly from the bound-state kinematics. In the non-relativistic limit, the top and antitop quarks have indeed small momentum in the $t\bar t$ rest frame, so their decay products tend to be correlated in angle. This leads to a higher probability for the lepton from the semi-leptonic top decay and one of the jets from the hadronic top decay to be close in the transverse plane. As a proxy for such an observable since we have no access to the flavour of the final-state light jets and whether they stem from a top decay, we rely on the $\min R^2_{\ell j}$ variable defined as
\begin{equation}
  \min R^2_{\ell j} = \min \Big(R^2_{\ell j_1}, R^2_{\ell j_2}\Big)\,,
\end{equation} 
where $j_1$ and $j_2$ denote the two light jets selected as originating from the decay of the hadronic top candidate, and $R^2_{\ell,j_1}$ and $R^2_{\ell,j_2}$ represent the square of the distance in the transverse plane between the lepton candidate and each of the two jets. The corresponding distribution for the signal (teal) and the conventional $t\bar t$ background (orange) is shown in figure~\ref{fig:R2lj}, both before (solid) and after (dashed) the selection on the reconstructed top-antitop invariant mass. The signal features a larger fraction of events at small separations, while the background is comparatively flatter and extends to larger values. This motivates the requirement
\begin{equation}\label{eq:cutR2}
  \min R^2_{\ell j} \leq 0.5\,.
\end{equation} 
After this selection, we obtain $N_\mathrm{toponium} = 3060$ signal events and $N_{t\bar t} = 81600$ background events. The corresponding significance $s$ and signal-to-noise ratio correspond then 
\begin{equation}\begin{split}
  s = \frac{N_\mathrm{toponium}}{\sqrt{N_\mathrm{toponium} + N_{t\bar t}}} = 10.5\,, \quad
\frac{N_\mathrm{toponium}}{N_{t\bar t}} = 3.75\%\,.
\end{split}\end{equation}

These results demonstrate that near-threshold toponium formation leaves a clear and characteristic imprint on single-leptonic $t\bar t$ events at the LHC. In particular, the angular correlations between the lepton and the jets from the hadronic top decay emerge as powerful discriminants with genuine sensitivity to bound-state effects. Even under the conservative assumptions of perfect reconstruction and no systematic uncertainties, we indeed find a statistical significance of \mbox{$s\!\sim\!11$} after a selection on one of such variables, which suggests that toponium effects could already be probed with the existing LHC Run~2 dataset. Furthermore, additional sensitivity may be gained by combining the implemented selection with other observables.

\begin{figure}
  \centering
  \includegraphics[width=\linewidth]{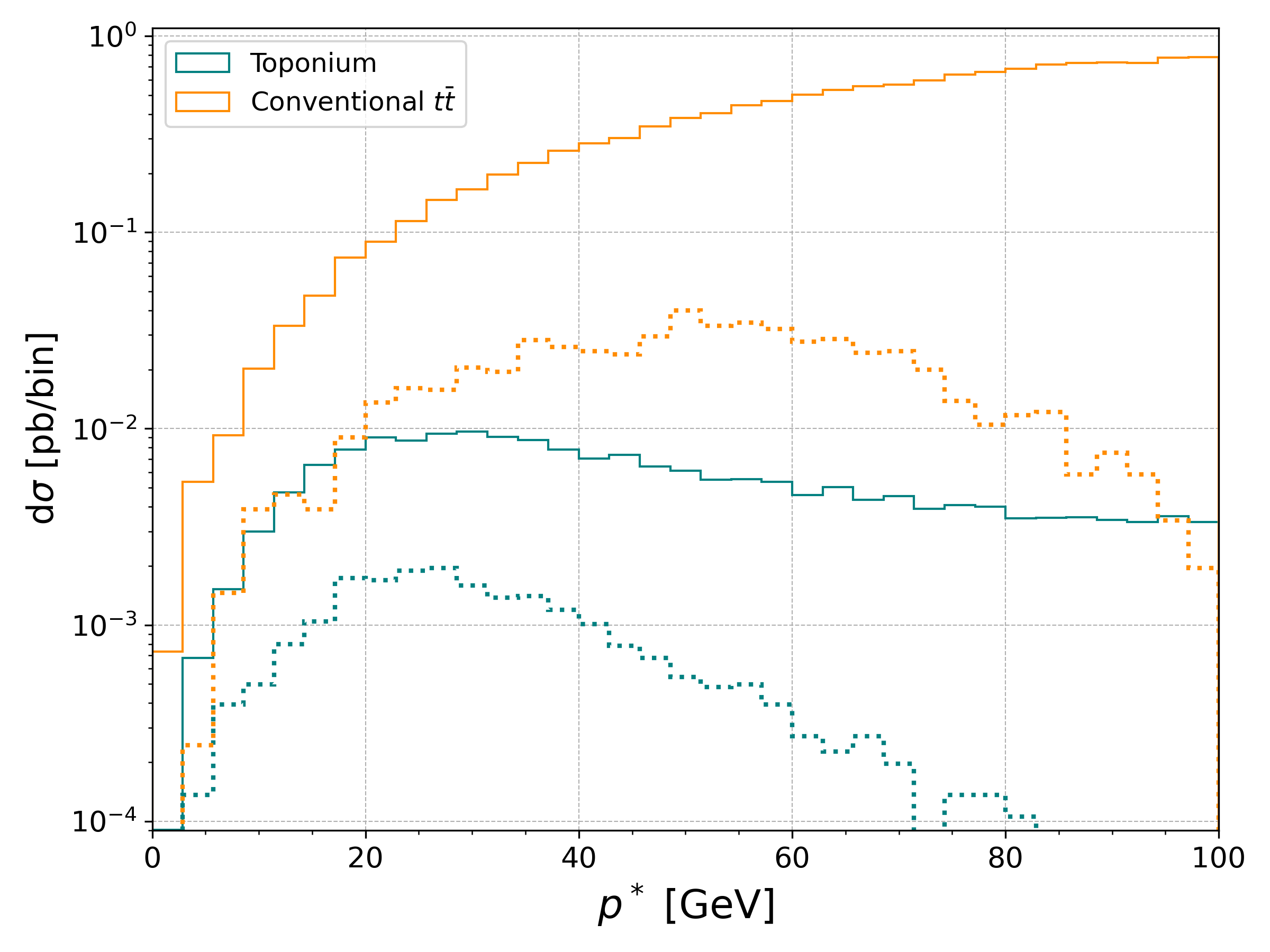}
  \caption{Differential cross section for toponium (teal) and conventional top-antitop (orange) production in proton-proton collisions at $\sqrt{s}=13$~TeV in the single-leptonic channel, shown as a function of the momentum $p^*$ of the reconstructed top quarks in the toponium rest frame. The predictions include a fiducial selection requiring exactly one lepton, two light jets, two $b$-jets and $\met > 30$~GeV (solid), as well as additional cuts in which the reconstructed invariant mass of the top-antitop system is constrained to be below 350~GeV and $\min R^2_{\ell j} \leq 0.5$ (dashed). \label{fig:p*}}
\end{figure}
After the above selection, we determine the momenta of the two top quarks in the $t\bar t$ rest frame and study the distribution of their common three-momentum magnitude $p^*$. The resulting spectra for the signal (teal) and the background (orange) are shown in figure~\ref{fig:p*}, both with only our basic selection of one lepton, two $b$-jets, two light jets and $\met > 30$~GeV (solid) and after enforcing first the reconstructed $t\bar t$ invariant mass to be smaller than 350~GeV and next the cut of eq.~\eqref{eq:cutR2} (dashed). In both cases, the signal and background distributions display markedly different behaviours: the toponium contribution peaks at \mbox{$p^* \sim 20$~GeV}, whereas the background rises towards larger values of the top momentum $p^*$. This variable therefore offers strong discriminating power and could be leveraged to increase the signal significance and the signal-to-noise ratio. It should however be stressed that genuine toponium events are not expected to populate the large-$p^*$ region since generator-level cuts of $m_{t\bar t} < 350$~GeV and $p^*<50$~GeV have been imposed in order to restrict the generated events to be in the near-threshold, non-relativistic regime where our methodology relying on the NRQCD-inspired re-weighting of the matrix elements is reliable (see section~\ref{sec:technics} and~\cite{Fuks:2024yjj}). Events with large reconstructed recoil momentum therefore predominantly arise from parton showering and reconstruction effects. The additional cut of eq.~\eqref{eq:cutR2} is thus doubly beneficial: it suppresses background contributions and enhances the significance of the toponium signal while simultaneously restricting the selected signal sample to events with reliable kinematic reconstruction.

Beyond its phenomenological utility, the top momentum $p^*$ in the toponium rest frame has a direct theoretical interpretation: it controls the structure of the toponium properties as a bound state, as predicted in the NRQCD framework where it enters the Green's function of the non-relativistic QCD Hamiltonian together with the binding energy $E$. The peak at \mbox{$p^* \approx 20$~GeV} is precisely the feature expected from theoretical calculations~\cite{Sumino:2010bv} and corresponds to the inverse Bohr radius of the top quark. Remarkably, this structure, already observed at the parton level in the dileptonic case~\cite{Fuks:2024yjj}, survives parton showering and full event reconstruction in the single-leptonic final state. This makes the momentum $p^*$ a key observable to investigate, so that measuring its distribution in data would provide a direct experimental handle on the bound-state dynamics of the top quark. It consequently represents the most promising avenue for characterising toponium formation at the LHC.

\section{Conclusion}\label{sec:conclusion}
In this letter, we have presented a dedicated study of near-threshold toponium formation effects in the single-leptonic channel at the LHC. Building on our recent framework integrating bound-state effects in standard Monte Carlo simulations for hadron colliders via the re-weighting of conventional matrix elements, we have extended the method to toponium events featuring a single lepton in the final state, thereby enabling simulations of toponium production and decay matched with parton showering in this channel. Our work therefore provides, for the first time, a practical approach to treat toponium bound-state effects in the single-leptonic mode by means of state-of-the-art Monte Carlo simulations.  

We have further employed our approach to demonstrate that toponium formation leaves characteristic imprints on kinematic observables in single-leptonic top-antitop events. In particular, angular correlations between the lepton and the hadronic top decay products already provide substantial discriminating power, and we have identified that the reconstructed momentum $p^*$ of the top quarks in the $t\bar t$ rest frame can further be studied as one of the most sensitive observable to toponium effects. This quantity not only offers strong separation between the toponium signal, defined as the set of selected events in which the intermediate $t\bar t$ system is in a colour-singlet state and with a re-weighting using the NRQCD Green's function, and the background embedding conventional perturbative contributions including double-resonant, single-resonant and non-resonant components as well as both the colour-singlet and colour-octet contributions. It also retains the imprint of the non-perturbative bound-state dynamics in its distribution, even after parton showering and event reconstruction. Remarkably, the peak expected from NRQCD at $p^*\sim 20$~GeV persists, and our estimates further suggest that a statistically significant signal could already be accessible in the LHC Run~2 dataset.

These findings highlight that the single-leptonic channel is a competitive and complementary avenue to the dileptonic mode for probing toponium formation at the LHC. By virtue of the larger statistics and the easier reconstruction of the final state, a measurement of the top recoil momentum $p^*$ distribution is not only in principle achievable, but could also provide a unique experimental handle on the bound-state dynamics of the top quark. Looking ahead, it will be important to refine these predictions and to combine, within the analysis, its study with that of other complementary observables. Together, these efforts could finally contribute to the characterisation of toponium and therefore its discovery, and bring it within reach of present or near-future data.

\section*{Acknowledgements}  
KM would like to thank Prof.~Shoji Hashimoto for his kind invitation to visit KEK. The work of BF was supported in part by Grant ANR-21-CE31-0013 (Project DMwithLLPatLHC) from the French \emph{Agence Nationale de la Recherche} (ANR), the one of YJZ by the JSPS KAKENHI Grant No.~23K03403 and the one of KM by a Shaanxi Fundamental Science Research Project for Mathematics and Physics (Grant No.25JSY031).

\bibliographystyle{JHEP}
\bibliography{toponium}
\end{document}